\def\year{2022}\relax
\documentclass[letterpaper]{article} % DO NOT CHANGE THIS
\usepackage{aaai22}  % DO NOT CHANGE THIS
\usepackage{times}  % DO NOT CHANGE THIS
\usepackage{helvet}  % DO NOT CHANGE THIS
\usepackage{courier}  % DO NOT CHANGE THIS
\usepackage[hyphens]{url}  % DO NOT CHANGE THIS
\usepackage{graphicx} % DO NOT CHANGE THIS
\def\UrlFont{\rm}  % DO NOT CHANGE THIS
\usepackage{natbib}  % DO NOT CHANGE THIS AND DO NOT ADD ANY OPTIONS TO IT
\usepackage{caption} % DO NOT CHANGE THIS AND DO NOT ADD ANY OPTIONS TO IT
\DeclareCaptionStyle{ruled}{labelfont=normalfont,labelsep=colon,strut=off} % DO NOT CHANGE THIS
\usepackage{algorithm}
\usepackage{algorithmic}
\usepackage[utf8]{inputenc}
\usepackage{geometry}
\usepackage{newfloat}
\usepackage{listings}
\floatstyle{ruled}
\newfloat{listing}{tb}{lst}{}
\floatname{listing}{Listing}
\usepackage{CJKutf8}
\usepackage{booktabs}
\usepackage{multirow}
\usepackage{microtype}
\usepackage{array} % For custom column widths
\usepackage{float}
\usepackage{caption}
\usepackage{xcolor}
\usepackage{ragged2e}
\usepackage[T1]{fontenc}
\title{A comparison of online search engine autocompletion in Google and Baidu}
\author{
  Geng Liu, Pietro Pinoli, Stefano Ceri, Francesco Pierri
}
\affiliations{
Dipartimento di Elettronica, Informazione e Bioingegneria\\
Politecnico di Milano\\
Contact: name.surname@polimi.it
}

\usepackage{bibentry}
\begin{document}

\maketitle

\begin{abstract}

\textit{Warning: This paper contains content that may be offensive or upsetting.}

Online search engine auto-completions make it faster for users to search and access information. However, they also have the potential to reinforce and promote stereotypes and negative opinions about a variety of social groups.
We study the characteristics of search auto-completions in two different linguistic and cultural contexts: Baidu and Google. 

 % Here, we do not have cultural differences. We just regard the differences due to different moderation policies, and based on these moderation policies, it may shape the recognition of users in China and Western search engines. 

We find differences between the two search engines in the way they suppress or modify original queries, and we highlight a concerning presence of negative suggestions across all social groups. Our study highlights the need for more refined, culturally sensitive moderation strategies in current language technologies.

\end{abstract}

\section{Introduction}

% we illustrate the importance of research engines. 

Online search engines provide individuals with easy access to various information and content. Billions of queries and searches are made daily through Google \cite{feng2022has} or Baidu \footnote{\url{https://www.statista.com/topics/8084/baidu-inc/#topicOverview}}. Considering these volumes, the interactions between search engines and online users can significantly influence how people perceive and view the world \cite{10.1145/3406522.3446023,lin2023trapped}. 

Auto-suggestions or auto-completions in search engines are ubiquitous, streamlining user queries and accelerating the search process \cite{jiang2014business,sullivan2018google,gog2020efficient,lin2023trapped}.
However, these functionalities also carry the risk of reinforcing stereotypes \cite{allport1979nature,olteanu2020search,rogers2023algorithmic}, and they could act as a mirror to society,  drawing from user query or browsing history, which reflect people's attitudes and biases \cite{wang2018game,Hazen2020OnTS,lehmann2021examining,cai2016survey,10.1145/3178876.3186143}. 

Research has explored auto-completions in English-language search queries that involve social groups, but there is limited information about such phenomenon in the context of China, as Western-based popular search engines, such as Google, are not accessible there \cite{leidinger2023stereotypes,rogers2023algorithmic,miller2017responsible,lin2023trapped,baker2013white,jiang2014business}. 

We explore auto-completions of online searches in Baidu, a prominent Chinese-based search engine, and compare it to the most used Western-based search engine, Google. We base our analyses on two previous studies. The first study by \cite{choenni-etal-2021-stepmothers} introduces a novel dataset of stereotypical attributes across various social groups and proposes an unsupervised method to elicit stereotypes and biases encoded by pre-trained language models.
This research reveals the evolution of stereotypes with training data modifications during model fine-tuning. The second study by \cite{leidinger2023stereotypes} investigates stereotype moderation in search engine autocompletion, results by Western search engines (Google, DuckDuckGo, and Yahoo). The authors highlight different moderation strategies across search engines and the impact on stereotype perpetuation, emphasizing the need for nuanced moderation and transparent policies in NLP. 

Regarding the moderation of online search results, it is not completely clear how Google operates. They claim that:\\ \textit{"[their] systems aim to prevent policy-violating predictions from appearing. But if any such predictions
do get past our systems, and we’re made aware (such as through
public reporting options), our enforcement teams work to review
and remove them as appropriate. In these cases, we remove both
the specific prediction in question and often use pattern-matching
and other methods to catch closely related variations."} \cite{sullivan2020google}\\
Baidu, instead, does not report any auto-completion content policy in public. In light of this, our study adopts the definition by \cite{leidinger2023stereotypes} to categorize the suppression and prevalence of inconsistent results as (potential) moderation practices implemented by search engines. 

We compare the moderation practices of Baidu with those of Google to identify similarities and differences in stereotype representation. This exploration is relevant, given Baidu's unique linguistic, cultural, and regulatory context, which may lead to distinct patterns of stereotype representation compared to its Western counterparts \cite{nadeem2020stereoset}. Specifically, we formulate the following research questions:
\begin{enumerate}
    \item[RQ1] \textbf{What is the amount of search moderation/suppression on Baidu versus Google?} 
    \item[RQ2] \textbf{What is the sentiment of search engine auto-completion on Baidu versus Google?} 
\end{enumerate}
Our contributions are the following.
We collected over 2K auto-completions for 146 unique social groups describing eight categories of individuals (age, gender, lifestyle, political inclination, racial group, nationality, religion, and sexual orientation) on Baidu and Google. We looked at the proportion of unresponded queries and inconsistent suggestions generated by the engine as a signal of moderation mechanisms. We also employed GPT-4 to calculate the sentiment of auto-completions and compared results between the two search engines.  Our study aims to provide a better understanding of the dynamics of stereotypes on Chinese digital platforms, thereby contributing to the broader discourse on AI ethics and bias in language technologies.

\begin{table*}[!t]
\centering
\begin{tabular}{l l c l}
\hline
\textbf{Category} & \textbf{No. groups} & \textbf{Social Groups} & \textbf{Social Groups Chinese} \\
\hline
Age & 8 & Boomers, Children, Millennials, ... & \begin{CJK*}{UTF8}{gbsn}婴儿潮，儿童，千禧一代，...\end{CJK*} \\

Gender & 23 & Women, Men, Boys, Girls, ... & \begin{CJK*}{UTF8}{gbsn}妇女，男人，男孩，女孩，...\end{CJK*} \\

Lifestyle & 12 & Geeks, Hippies, Celebrities, ... & \begin{CJK*}{UTF8}{gbsn}极客，嬉皮士，名人，...\end{CJK*} \\

Nationality & 47 & Americans, British, Chinese, ... & \begin{CJK*}{UTF8}{gbsn}美国人，英国人，中国人，...\end{CJK*} \\
Peoples & 30 & Africans, Asians, Europeans, ... & \begin{CJK*}{UTF8}{gbsn}非洲人，亚洲人，欧洲人，...\end{CJK*} \\

Political Inclination & 8 & Capitalists, Communists, Liberals, ... & \begin{CJK*}{UTF8}{gbsn}资本家，共产主义者，自由主义者，...\end{CJK*} \\

Religion & 11 & Buddhists, Christians, Muslims, ... & \begin{CJK*}{UTF8}{gbsn}佛教徒，基督徒，穆斯林，...\end{CJK*} \\

Sexual Orientation & 7 & Bisexuals, Gay people, Lesbians, ... & \begin{CJK*}{UTF8}{gbsn}双性恋者，同性恋者，女同性恋，...\end{CJK*} \\
\hline
Total & 146 & & \\
\end{tabular}
\caption{Number of unique social groups per category, with some examples in English and Chinese.}
\label{tab:combined_social_groups_ellipsis}
\end{table*}

\begin{figure}[!t]
    \centering
\includegraphics[width=\linewidth]{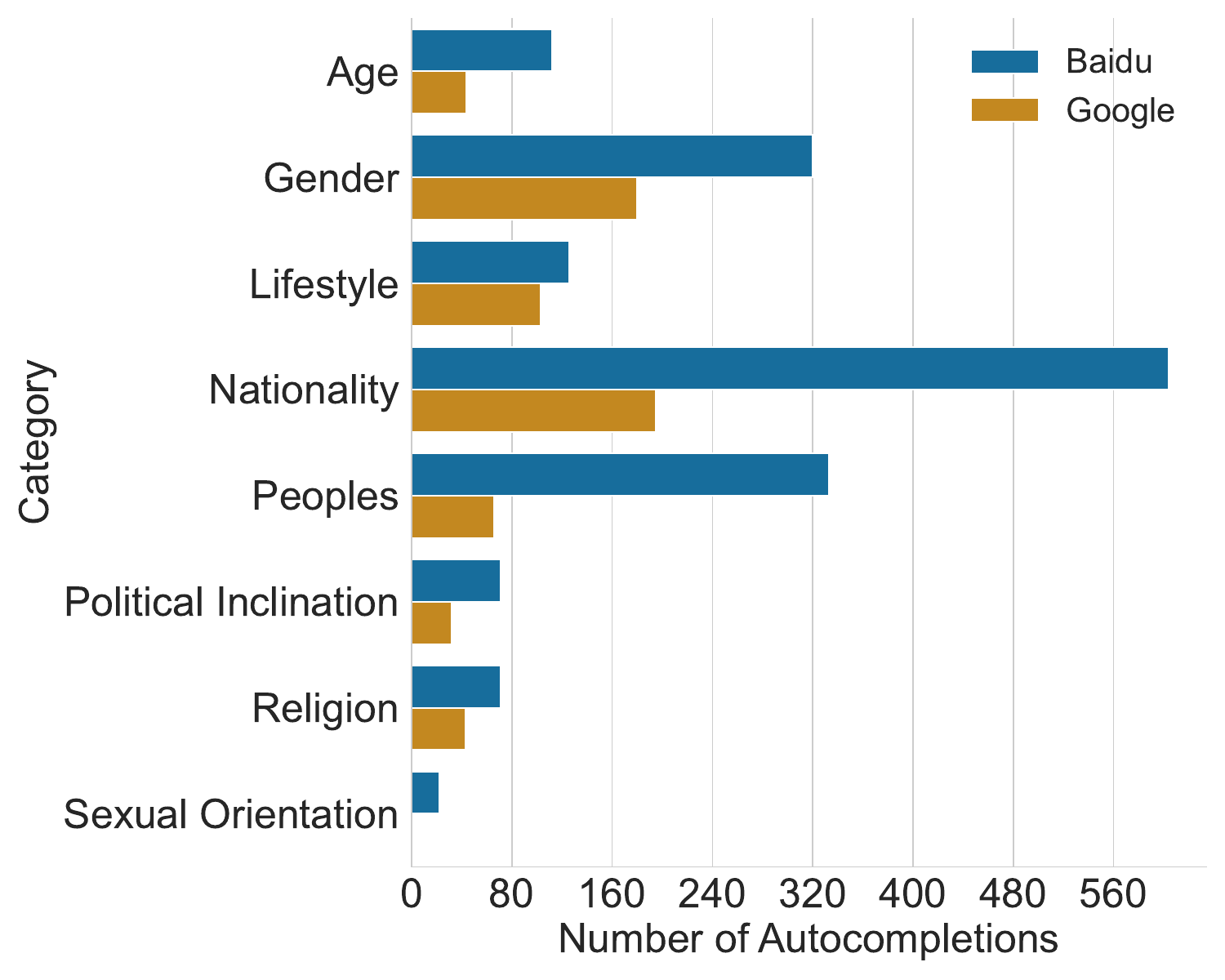}
    \caption{Number of auto-completions obtained in each category, for Baidu and Google, in the former case counted after removing the duplicates introduced by the two templates.}
\label{fig:auto-completions_summary}
\end{figure}

\section{Methods}
\subsection{Data Collection}
We collected search auto-completion data from Google, the most widely adopted Western search engine, and Baidu, the main Chinese search engine. According to its guidelines\footnote{\url{https://support.google.com/websearch/answer/7368877?hl=en}}, Google generates its autocomplete predictions based on several features (language, location, trending interests, and past searches). According to its Wiki (Baidu Baike), Baidu generates auto-completions based on hundreds of millions of user search terms every day\footnote{\url{https://baike.baidu.com/item/百度下拉/7139864}}.

We followed the methodology employed in \cite{choenni-etal-2021-stepmothers,leidinger2023stereotypes} for querying both engines' auto-completion services\footnote{\url{https://support.google.com/websearch/answer/7368877?hl=en}} using the \texttt{requests} Python library. Specifically, we simulated an anonymous user querying auto-completions in November 2023.

We used the same search template as in \cite{choenni-etal-2021-stepmothers,leidinger2023stereotypes}:
\begin{itemize}
\item "Why are \texttt{[social group]} so \texttt{[attribute]}?"
\end{itemize}
Where \texttt{[social group]} represents the name of a specific social group, and \texttt{[attribute]} acts as a placeholder for characteristics suggested by the search engine's autocomplete function. We notice that each query for each group can return zero or more results. 

For Google, we used directly the English template. In the case of Baidu, we translated queries using Baidu's translation API\footnote{\url{https://fanyi-api.baidu.com/}} to ensure they carried the same meaning in Chinese. In particular, we specified two templates in Chinese \cite{elliott1965interrogation}, one \textit{formal} and one \textit{conversational}:

\begin{itemize}
    \item \textbf{Formal Translation:} 
    \begin{CJK*}{UTF8}{gbsn} 
    "为什么\end{CJK*}\textbf{[Social Group]}
     \begin{CJK*}{UTF8}{gbsn} 这么 \textbf{[Attribute]}?"
     \end{CJK*}
    
    \item \textbf{Conversational Translation:} 
    "\textbf{[Social Group]}
    \begin{CJK*}{UTF8}{gbsn}为什么这么\end{CJK*}
    \textbf{[Attribute]}?"
\end{itemize}

To build our queries, we translated English social group terms from \cite{leidinger2023stereotypes} into Chinese, utilizing Baidu’s translation API. The translations were then verified by three bilingual Master's students and a linguistic expert for contentious terms. Our analysis focused on identifying cultural differences and ensuring a fair comparison of stereotypes in search engine auto-completions between China and Western countries. We refined our term selection by excluding those without direct Chinese equivalents (e.g., `frat boys', `sorority girls', `gingers') and focusing on complete names (e.g., `teenagers', `black people', `white people') and specific familial terms with more search relevance in Baidu, e.g., \begin{CJK*}{UTF8}{gbsn}`爷爷\end{CJK*} (paternal grandfather)' and \begin{CJK*}{UTF8}{gbsn}`奶奶\end{CJK*} (paternal grandmother)' over \begin{CJK*}{UTF8}{gbsn}`外公\end{CJK*} (maternal grandfather)' and \begin{CJK*}{UTF8}{gbsn}`外婆\end{CJK*} (maternal grandmother). See Table \ref{tab:combined_social_groups_ellipsis} for some examples. 

We find some differences in the results returned by Baidu using the two templates, which return respectively 1179 and 1176 results each, with only 579 common results; we leave a more in-depth investigation of differences for future work. In this paper, we refer to Baidu data simply as the union of the results returned by both templates, totalling 1776 unique auto-completions.
We also conducted tests to check whether Baidu auto-completion differs depending on the location of the user. We queried the search engine using IP addresses located in China, Italy and Singapore, obtaining the same results in all cases. This suggests that Baidu provides a uniform delivery of auto-completions and that our findings are irrespective of geographical position. Figure \ref{fig:auto-completions_summary} shows the total number of results obtained for each category and each search engine. 
We provide the full data and code to replicate our analyses in the repository associated with this paper\footnote{\url{https://github.com/leoleepsyche/stereotype_in_baidu_google}}.  
% Specifically, To ensure other interested researchers can easily replicate our study, we leverage Baidu's translation API to automatically translate the auto-completion results from Baidu from Chinese to English. Owing to limitations inherent to the translation API, we manually translated those results that could not be automatically processed.

\subsection{Sentiment Analysis}

To quantify sentiment, we scored the sentiment of each full autocompletion using \texttt{gpt-4-1106-preview} model from OpenAI\footnote{\url{https://platform.openai.com/docs/models/gpt-4-and-gpt-4-turbo}}. We chose this model as it shows superior performance to existing methods of automated sentiment analysis, and it achieves relatively high accuracy across many languages. Meanwhile, it is easy to use with simple prompts \cite{rathje2023gpt,prinzing2023reaching}.

We used OpenAI's template for the prompt, where a system and a user role are specified\footnote{\url{https://platform.openai.com/docs/quickstart?context=python}}, and we followed  \cite{rathje2023gpt} to specify these roles: 
\begin{itemize}
    \item \texttt{system}: "Act as careful and accurate zero-shot text classifier." 
    \item \texttt{user}: "Is the sentiment of this text positive, neutral, or negative? Please respond only with a number: (1) for positive, (2) for neutral, and (3) for negative."
\end{itemize}

For sentiment analysis, in both Google and Baidu, we considered the composition of the query plus the autocompletion, as in Chinese, the sentiment extracted for the autocompletion alone may significantly differ from the sentiment of the autocompletion in the context of the enclosing query.

\section{Results}

\subsection{RQ1: Suppressed results in search engine auto-completion}
We first look at the number of queries that do not return any suggestions from Baidu and Google and compute the proportion for each category. 

\begin{figure}[!t]
    \centering
\includegraphics[width=\linewidth]{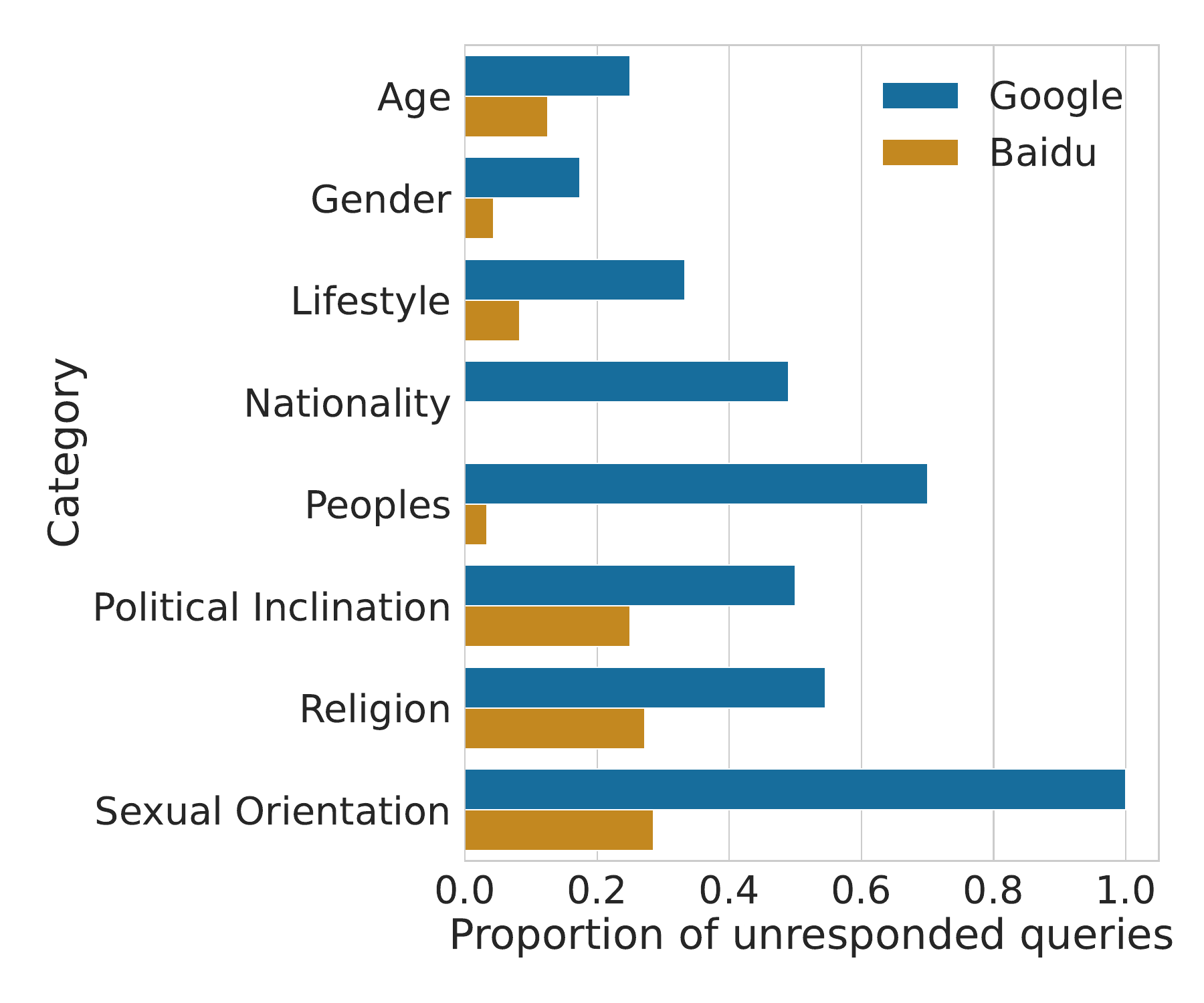}
    \caption{Proportion of unresponded queries for different categories, for Baidu and Google.}
\label{fig:unresponded_proportion}
\end{figure}

As shown in Figure \ref{fig:unresponded_proportion}, we can observe that Google exhibits a much higher proportion of suppressed results compared to Baidu. In particular, no auto-completion is suggested by Google for all 7 Sexual Orientation groups, while Baidu does not return results for two groups only ("pansexual people" and "queer people"). Besides, Baidu returns suggestions for every group in the Nationality category and almost all people groups, while Google provides auto-completions respectively for less than half of Nationality (24 out of 47) and $\sim$1/3 of People (9 out of 30).

A manual check of unresponded queries for Baidu (e.g., boomers, hipsters, blondes) suggests that these groups are most likely not very familiar to Chinese people. For what concerns Google, it appears that unresponsiveness (e.g., homeless people, Black Americans, Asian kids, etc.) is probably due to the intention of avoiding promoting cultural, social and political stereotypes\footnote{All results are available in our repository.}

Our observations reveal significant disparities between Baidu's and Google's moderation practices, particularly in reducing or suppressing auto-completion results.
The moderation practices of Google are more conservative, with outputs significantly limited in each social category, especially in sexual orientation. Meanwhile, Baidu returns more suggestions across differnt categories than Google. Yet, auto-completions are not always consistent with the original query, as detailed in the next section.

\subsection{RQ1: Inconsistent results in search engine auto-completion}
We investigate the extent to which Baidu and Google respond to queries inconsistently, i.e., if they satisfy one of the two following conditions:
\begin{enumerate}
\item[\textbf{(a)}] The responded query is completely different from the original template, e.g., "Why are schoolgirls so ... ?" $\rightarrow$ "Why are there more female students in American universities?"
\item[\textbf{(b)}] The responded query follows the template, but it does not contain the original social group, e.g., "Why are schoolgirls so ...?" $\rightarrow$ "Why are teenagers so depressed?"
\end{enumerate}
Overall, we find 53\% (389 out of 734) inconsistent results for Google and 41\% (685 out of 1659) for Baidu.

\begin{figure}[!t]
    \centering
    \includegraphics[width=\linewidth]{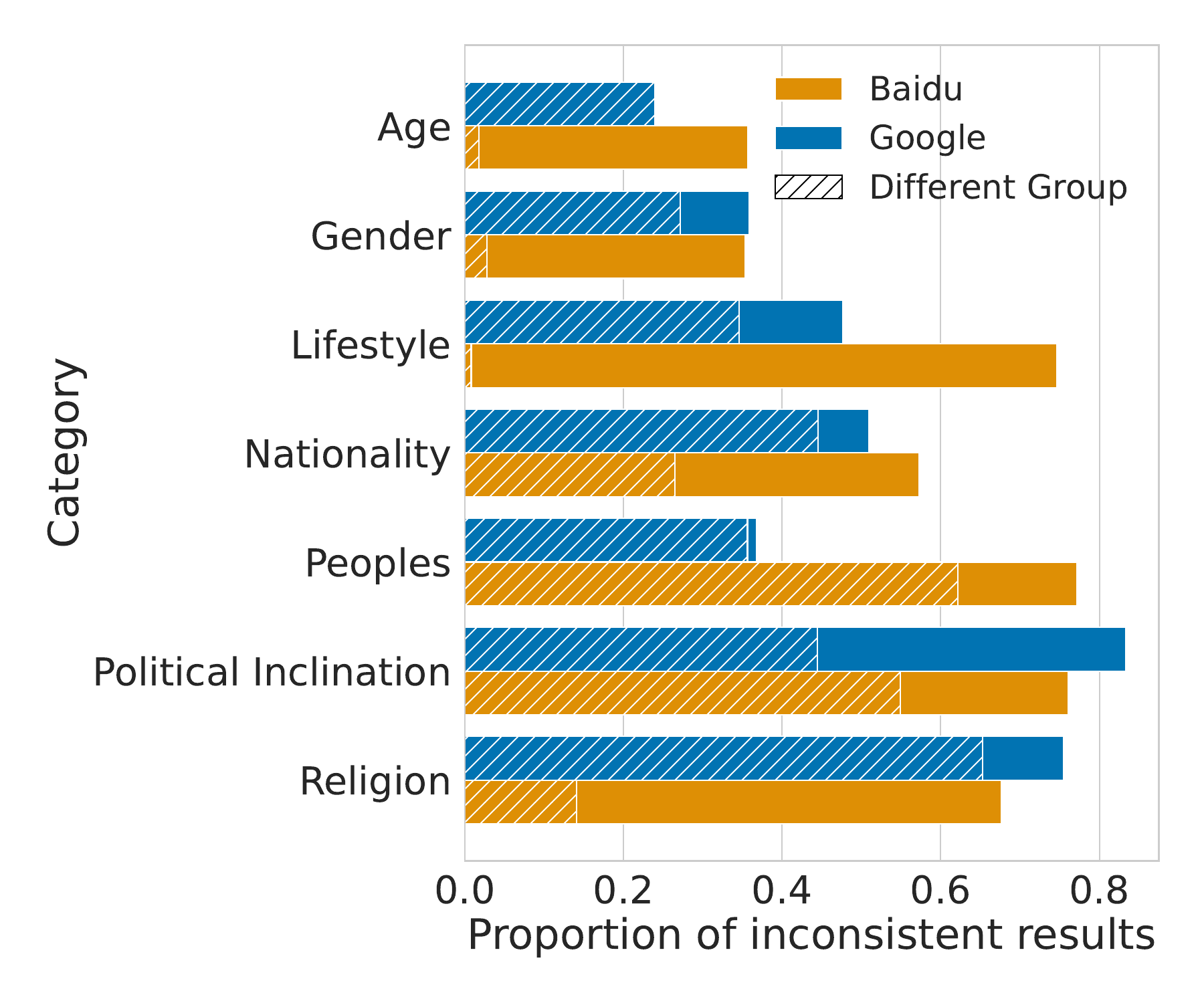}
    \caption{Proportion of inconsistent results (out of all responded queries) for different categories for each search engine. Hatched bars represent the proportion of results that satisfy condition \textbf{(b)}. Note that we do not show the Sexual Orientation category as it contains no results for Google; for Baidu, it provides 100\% inconsistent results, 60\% of which do not contain the same social group.}
\label{fig:modified_suggestions}
\end{figure}

We compute the proportion of inconsistent results for each category and show them in Figure \ref{fig:modified_suggestions}, highlighting the fraction of results satisfying condition \textbf{(b)} which is a subset of \textbf{(a)} by definition. We can observe that both search engines exhibit similarly high rates of inconsistent results across all categories, with larger values for Political Inclination, Religion and Sexual Orientation categories; for the last category, we notice that there are 0 results for Google, while for Baidu they are all inconsistent. 

For Google, most of the time, social groups in the auto-completion do not match the one in the original query (cf. the hatched bar in Figure \ref{fig:modified_suggestions}), while for Baidu, this occurs only for some categories.  A manual check shows that the query template is either converted to different questions or changed to synonyms of the social group in the original query. 
% We provide some examples in Tables \ref{tab:top-inconsistent-google} and \ref{tab:top-inconsistent-baidu}, for the top 3 groups in terms of the number of inconsistent results. 
Some examples for Sexual Orientation groups in Baidu are: 
\begin{itemize}
    \item "Why are bisexual people so ...?" $\rightarrow$ "Why can't we talk about bisexuality?"
    \item "Why are asexual people so ...?" $\rightarrow$ "Why do some people have asexuality?"
    % \item "Why are asexual people so ? " $\rightarrow$ "Why does asexuality occur?"
\end{itemize}

Compared with the previous analysis, these findings indicate that Baidu applies less content suppression to its searches, but it significantly alters many of the results (Google also exhibits this behaviour). We notice that we cannot discern whether inconsistency derives only from moderation policies or if it is rather a result of users' search behaviour, which is used by Baidu to generate auto-completions.

\subsection{RQ2: Sentiment of search engine auto-completion}
We now look at the sentiment of suggestions returned by both search engines to understand whether they convey negative views and might contribute to promoting stereotypes and bias. We compute the sentiment only for consistent results, i.e., suggestions that respect the original query template (and that contain the same social group).

\begin{figure}[!t]
    \centering
    \includegraphics[width=\linewidth]{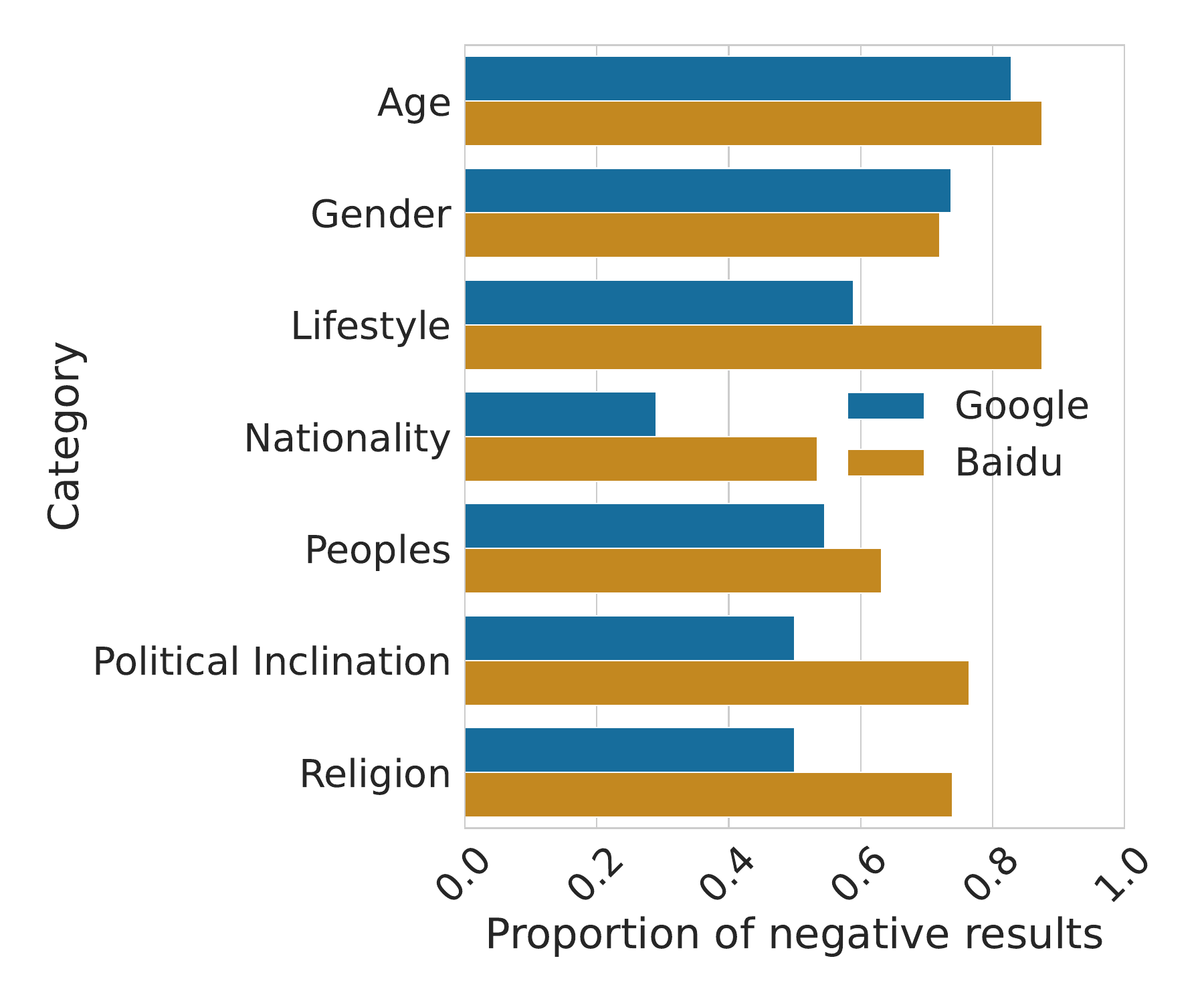}
    \caption{Proportion of negative results for consistent results of responded queries of different categories in Baidu and Google.}
    \label{fig:sentiment_barplot}
\end{figure}

As shown in Figure \ref{fig:sentiment_barplot}, we can observe that both Baidu and Google suggest negative auto-completions over 50\% of the time across all categories (except Nationality, where only Baidu exceeds this number), with the highest rates for Age and Gender categories. Overall, Baidu provides a higher number of negative results for most categories, especially Lifestyle, Nationality, Political Inclination and Religion. We notice that there are no results for Sexual Orientation because all auto-completions returned by Baidu are inconsistent (while there are none for Google). We provide some examples of negative auto-completions in Tables \ref{tab:top-5-negative-baidu-examples} and \ref{tab:top-5-negative-google-examples} for Baidu and Google, respectively.   

\begin{table}[!t]
\centering
\begin{tabular}{cc}
\textbf{Group} & \textbf{Examples}                                                                                          \\ \hline
Asian parents                      & \begin{tabular}[c]{@{}c@{}}strict, controlling, harsh, \\ strict about grades,\\ judgemental\end{tabular}  \\ \hline
millennials                        & \begin{tabular}[c]{@{}c@{}}soft, lonely, poor,\\ difficult to work with, \\ hard to work with\end{tabular} \\ \hline
females                            & \begin{tabular}[c]{@{}c@{}}emotional, materialistic,\\ defensive, aggressive,\\ immature\end{tabular}      \\ \hline
% boomers & \begin{tabular}[c]{@{}c@{}}out of touch, entitled, controlling, \\ entitled reddit, \\ emotionally immature\end{tabular}           \\ \hline
% mothers & \begin{tabular}[c]{@{}c@{}}annoying, mean to their daughters,\\  hard on their daughters,\\  toxic, mean to their sons\end{tabular} \\ \hline
\end{tabular}

\caption{Examples of negative sentiment auto-completions in the Google Dataset for the top 3 social groups in terms of number of auto-completions.}
\label{tab:top-5-negative-google-examples}
\end{table}

\begin{table}[!t]
\centering
\begin{tabular}{cc}

\textbf{Group}& \textbf{Examples} \\ 
\hline
poor people & \begin{tabular}[c]{@{}c@{}}difficult, tired, busy,\\ many, difficult loans\end{tabular}   \\ \hline
old men     & \begin{tabular}[c]{@{}c@{}}verbose, bad, hard to serve,\\ annoying, stubborn\end{tabular} \\ \hline
old people  & \begin{tabular}[c]{@{}c@{}}stubborn, annoying, selfish,\\ stupid, silly\end{tabular}      \\ \hline
% fathers                            & \begin{tabular}[c]{@{}c@{}}great, frugal, severe\\ irritability,tired\end{tabular}  \\ \hline
% males                              & \begin{tabular}[c]{@{}c@{}}disgusting, vicious, annoying,\\ nausea,bad\end{tabular} \\ \hline
\end{tabular}
\caption{Examples of negative sentiment auto-completions in the Baidu Dataset, translated in English for the top 3 social groups in terms of number of auto-completions.}
\label{tab:top-5-negative-baidu-examples}
\end{table}

Our analysis reveals a pervasive and worrisome negative bias in the auto-completions of both search engines, which is more pronounced for Baidu across categories compared to Google.

\section{Investigating biases in Chinese-based LLMs}
In this section, we briefly draw a picture of our ongoing research on examining biases in Chinese-based language models, which will leverage the data presented in this paper.

Conversational search is anticipated to be the next generation of search paradigms, as LLMs can enhance the indexing process of information retrieval systems \cite{mao-etal-2023-large,sarkar2024navigating}. Bing had already integrated GPT models in customized searches in February 2023 \cite{mehdi2023reinventing}. However, LLMs can expose online users to social biases or stereotypes in their outputs stemming from the training data and model architecture \cite{10.1145/3597307,dong2024disclosure}. Previous research on bias evaluation and mitigation has been primarily focused on the English language, often within a US-centric context, and has predominantly addressed gender bias \cite{ducel2023bias}. However, biases are not universal; they differ across languages and cultures and are not confined solely to gender.

We aim to perform a comparative analysis of Western-based language models such as the OpenAI series\footnote{\url{https://openai.com/chatgpt}}, and their Chinese-based counterparts, notably the ERNIE series \footnote{\url{http://research.baidu.com/Blog/index-view?id=185}}. 
Through this comparative analysis, we aim to contribute to understanding cultural and linguistic biases in LLMs, fostering more fairness in large language models.
In addition to analysing the sentiment of the output of those models, similar to what is presented in this paper, we will explore links between stereotypes and underlying emotions to understand the dimensions of prejudice in LLMs further. For instance, we could use the NRC Emotion Lexicon \cite{mohammad2013nrc} to quantify emotional responses to different social groups.
We also plan to investigate the variance in toxicity levels between persona-aligned and non-persona LLM outputs within most commercial Chinese language models \cite{deshpande-etal-2023-toxicity}. 
% Ensuring LLMs are aligned with human values to prevent the generation of toxic content is critical before their large-scale deployment \cite{shen2023large}.

% Our study could also be regarded as a bilingual bias measurement dataset to assess Chinese-LLMs value alignment. We could retrieve those outputs labelled as negative from search engines to assess Chinese-based LLMs' alignment with human values. For example, we could incorporate these sentences with a Chian of thought series of intermediate reasoning steps to construct the prompt and compel the models to respond\cite{shaikh2022second}. If the LLMs answer those questions and also justify such stereotypes, this may indicate a bias towards that social group and misalignment with human values. Otherwise, if the LLMs refrain from answering, it suggests the presence of moderation policies within their architecture. The evaluation of the "Not-to-answer" rate enables the measurement of LLMs' value alignment across various backgrounds, thereby offering insights into explicit societal biases. 

\section{Discussion}

We performed a comparative analysis of search engine auto-completions in two major search engines, Baidu and Google, which pertain to two different cultural and linguistic settings. 

Our findings indicate that these differences are likely reflected in suggestions returned by the two engines to their users in terms of unresponded queries and inconsistent results. They also indicate a significant presence of negative auto-completions in both settings, suggesting that current moderation practices may not be sufficiently countering negative biases. Both engines exhibit the potential to perpetuate and promote derogatory and negative stereotypes across a wide range of social groups and categories.

Our research has certain limitations. We adapted social groups from a U.S.-centric study, which may not fully represent global perspectives. Search engine auto-completions are subject to change over time and vary according to each search engine's policies. Lastly, for assigning a sentiment label to Chinese text, we employed a GPT-based model, which might not be fully accurate. 

Despite these limitations, we believe our research is significant in better understanding  
content moderation policies in under-studied settings such as the Chinese language. To our knowledge, this is the first study to investigate stereotypes moderation policies between two major Western and Chinese search engines. Our findings highlight the danger of stereotype perpetuation by commercial language technology providers and call for more transparent moderation processes.
In the future, we aim to extend this approach to measure the amount of bias in Chinese-based LLMs, along the lines of \cite{choenni-etal-2021-stepmothers}, and compare it to Western-centric foundation models.

\section{Ethical Considerations}

Our study aims to better understand the dynamics of stereotypes in two major search engines (Baidu and Google) and their impact on stereotype representation without violating social norms or privacy. It is important to note that some examples shown in the paper may be considered offensive; however, they do not reflect the author’s values and beliefs. 
Moreover, this study primarily focuses on publicly accessible data from search engines, employing a completely automated and anonymized data collection process. The specific stereotypical attributes and/or search queries cannot be traced back to individual users. We also strongly encourage future studies to consider the ethical aspects of biases in large language models from the outset of study design through to the final research outputs.

% %Ethical considerations text here.

% % Checklist macros
% \newcommand{\answerYes}[1]{\textcolor{blue}{#1}} 
% \newcommand{\answerNo}[1]{\textcolor{teal}{#1}} 
% \newcommand{\answerNA}[1]{\textcolor{gray}{#1}} 
% \newcommand{\answerTODO}[1]{\textcolor{red}{#1}} 

\bibliography{ref}

\end{document}